\documentclass[twoside]{article}
\usepackage{qic,epsfig,cite}
\begin{document}
\setlength{\textheight}{8.0truein}    

\runninghead{Experimental and theoretical challenges for the trapped electron quantum
computer} {Marzoli et. al.}

\normalsize\textlineskip
\thispagestyle{empty}
\setcounter{page}{1}

\copyrightheading{0}{0}{2003}{000--000}

\vspace*{0.88truein}

\alphfootnote

\fpage{1}

\centerline{\bf EXPERIMENTAL AND THEORETICAL CHALLENGES} \vspace*{0.035truein}
\centerline{\bf FOR THE TRAPPED ELECTRON QUANTUM COMPUTER} \vspace*{0.37truein}
\centerline{\footnotesize IRENE MARZOLI} \vspace*{0.015truein}
\centerline{\footnotesize PAOLO TOMBESI} \vspace*{0.015truein}
\centerline{\footnotesize GIACOMO CIARAMICOLI} \vspace*{0.015truein}
\centerline{\footnotesize\it Dipartimento di Fisica, Universit\`a di Camerino}
\baselineskip=10pt \centerline{\footnotesize\it 62032 Camerino, Italy} \vspace*{10pt}
\centerline{\footnotesize G\"UNTER WERTH} \vspace*{0.015truein}
\centerline{\footnotesize PAVEL BUSHEV} \vspace*{0.015truein} \centerline{\footnotesize
STEFAN STAHL} \vspace*{0.015truein} \centerline{\footnotesize\it Dept. of Physics,
Johannes-Gutenberg-University Mainz} \baselineskip=10pt \centerline{\footnotesize\it
55099 Mainz, Germany} \vspace*{10pt} \centerline{\footnotesize FERDINAND SCHMIDT-KALER}
\vspace*{0.015truein} \centerline{\footnotesize MICHAEL HELLWIG} \vspace*{0.015truein}
\centerline{\footnotesize\it Quanten-Informationsverarbeitung, Universit\"at Ulm}
\baselineskip=10pt \centerline{\footnotesize\it 89069 Ulm, Germany} \vspace*{10pt}
\centerline{\footnotesize CARSTEN HENKEL} \vspace*{0.015truein}
\centerline{\footnotesize\it Institut f\"ur Physik und Astronomie, Universit\"at Potsdam} \baselineskip=10pt
\centerline{\footnotesize\it 14476 Potsdam, Germany} \vspace*{10pt} \centerline{\footnotesize
GERRIT MARX} \vspace*{0.015truein} \centerline{\footnotesize\it Ernst-Moritz-Arndt
Universit\"at Greifswald} \baselineskip=10pt \centerline{\footnotesize\it 17489
Greifswald, Germany} \vspace*{10pt} \centerline{\footnotesize IGOR JEX} \vspace*{0.015truein}
\centerline{\footnotesize\it Department of Physics, FJFI \v{C}VUT} \baselineskip=10pt
\centerline{\footnotesize\it 115 19 Praha, Czech Republic} \vspace*{10pt} \vspace*{10pt}
\centerline{\footnotesize EWA STACHOWSKA} \vspace*{0.015truein}
\centerline{\footnotesize GUSTAW SZAWIOLA} \vspace*{0.015truein}
\centerline{\footnotesize ADRIAN WALASZYK} \vspace*{0.015truein}
\centerline{\footnotesize\it Poznan University of Technology} \baselineskip=10pt
\centerline{\footnotesize\it 60-965 Poznan, Poland} \vspace*{0.225truein}
\publisher{(received date)}{(revised date)}

\vspace*{0.21truein}

\abstracts{
We discuss quantum information processing with trapped electrons. After recalling the
operation principle of planar Penning traps we sketch the experimental conditions to
load, cool and detect single electrons. Here we present a detailed investigation of a
scalable scheme including feasibility studies and the analysis of all important
elements, relevant for the experimental stage. On the theoretical side, we discuss
different methods to couple electron qubits. We estimate the relevant qubit coherence
times and draw implications for the experimental setting. A critical assessment of
quantum information processing with trapped electrons is concluding the article. }{}{}

\vspace*{10pt}

\vspace*{3pt}
\communicate{to be filled by the Editorial}

\vspace*{1pt}\textlineskip

\section{Introduction}

{\em Importance and applications of quantum computing: } Quantum computers (QC) are
known to perform certain computational tasks exponentially more efficiently than their
classical counterparts. The theoretical concept of QC \cite{CHUANG00} is highly
developed and especially well-known is the algorithm for the factorization of large
numbers \cite{SHO94,EKERT96} because it threatens the entire security of commonly used
encryption schemes \cite{RSA2,RSA1}. Furthermore, efficient quantum algorithms exist
for searching entries in an unsorted data base \cite{GRO97}. More recently,
the simulation of quantum spin systems \cite{LLO96,JAN03,DEN05,POR04,Ciaramicoli2008} has become a focus of
research. World-wide efforts aim at a scalable realization of a QC \cite{ARDAEU}.

{\em Experimental approaches: } Various schemes have been proposed in the past decade
to realize experimentally a first model of a QC and we have seen remarkable
experimental and theoretical advances. An overview of experimental techniques towards
quantum computing is found in \cite{CHUANG00,BRUSS2006,Bouwmeester2000}. At present,
experiments with trapped cold ion crystals confined in linear Paul traps may be
regarded as most advanced \cite{CIR95,HAF05,LEI05,LEI03,SCH03,SCH03a,WIN03}. The main
experimental roadblocks are decoherence and scalability. We are presenting a new
approach based on \cite{CIA03} trying to overcome both these problems. Single electrons
can be trapped in an array of Penning traps at 100~mK. Working in a cryogenic
environment with only static electric and magnetic fields serves for low decoherence
rates. In a two dimensional array of micro structured Penning traps each electron
stores quantum information in its internal spin and motional degrees of freedom. This
architecture allows scalability, similar as in optical lattice experiments but with
individual control of qubit sites.

{\em The paper is organized as follows:} After introducing the experimental setup of an
array of planar Penning traps, we discuss the coupling of two qubits. Taking into
account the experimental conditions we estimate single and two qubit gate times, and
compare these values with the theoretically expected coherence time of electrons in the
vicinity of the planar trap electrode. This reasoning allows us to give a critical
assessment of the trapped electron approach for a future scalable QC.

\section{Penning traps}
\subsection{Basic properties}
\begin{figure} [htbp]
    \centerline{\epsfig{file=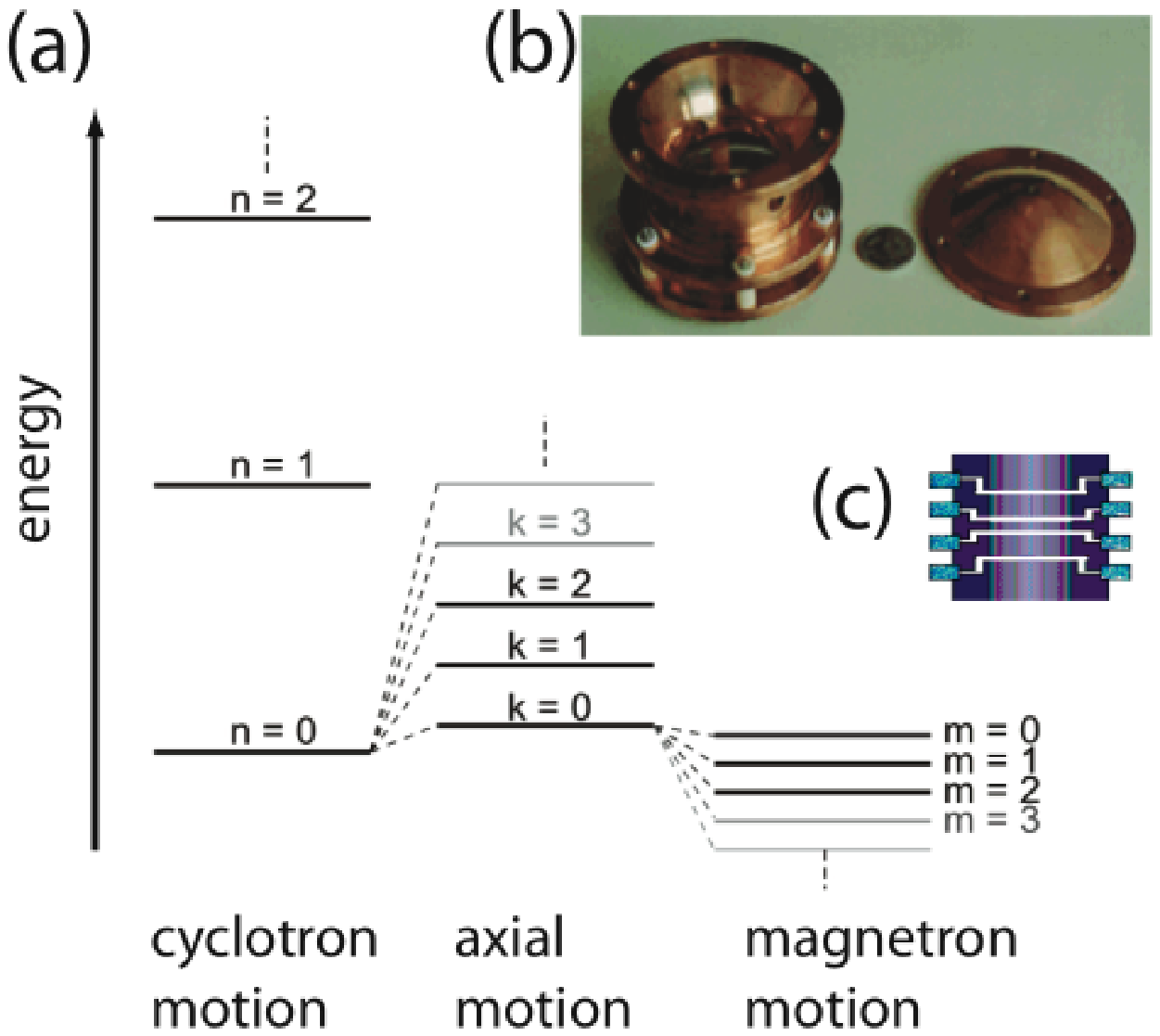, width=8cm}}
    \vspace*{13pt}
    \fcaption{\label{Fig:quantLevel} (a) Quantized energy levels of motion for
    an electron in a Penning trap. (b) Hyperbolic geometry fabricated in copper
    with cm dimensions for the ideal Penning trap and (c) side view of the design
    of trap electrodes out of a stack of rings.}
\end{figure}

The genuine three dimensional Penning trap \cite{BRO86,MAJ06} is composed of a
quadrupole electrical potential $\Phi^{el.}$ provided by a voltage $U$ applied between
a ring electrode and two electrically isolated end cap electrodes of hyperbolic 
shape [see Fig.~\ref{Fig:quantLevel}(b)],
and a superimposed constant magnetic field B$_0$ in direction of the $z$-axis
\begin{eqnarray}
  \Phi^{el.}(\rho,z) &=& \frac{U}{r_0^2} \left(\rho^2-2z^2 \right) .
\end{eqnarray}
The characteristic dimension of trap electrodes is denoted by $r_0$. For a properly
chosen polarity the electric field serves for confinement between the two end cap
electrodes in the axial direction (along the $z$-axis) while the magnetic field prevents
the electrons from escaping in the radial direction $\rho$. The
motion in this potential can be solved analytically and consists of three harmonic
oscillations at the frequencies
\begin{eqnarray}
  \omega_{\pm} &=& (\omega_c \pm \sqrt{\omega_c^2 - 2 \omega_z^2})/2 \\
   \omega_z &=& \sqrt{2eU/md^2}
\end{eqnarray}
where $\omega_c = e B_0 / m$ is the free electron cyclotron frequency, $\omega_+$ is
called the reduced cyclotron frequency, $\omega_-$ the magnetron frequency, and
$\omega_z$ the axial frequency. For our concept these frequencies are of the order of
$\omega_+/(2\pi)$ $\sim $100~GHz, $\omega_-/(2\pi)$ $\sim $10~kHz, $\omega_z/(2\pi)$
$\sim$ 100~MHz. We note that in the radial direction the electron moves on a potential
hill, making the magnetron motion metastable. This also shows up in a diagram of
quantized equidistant motional levels, see Fig.~\ref{Fig:quantLevel}, where the energy
of the magnetron motion decreases with increasing quantum number.

For the electron, the $g$-factor \cite{HANNEKE2008} determines the 
splitting of spin states
via the Larmor frequency $\omega_{L} = g e B_0 / 2 m$ which is slightly 
above, but
well separated from the electron motional frequency component $\omega_+$. Electron spin
states are very long lived such that we can identify the qubit computational basis states
of the quantum register with the spin state of each electron $\{|0\rangle \equiv
\left|\downarrow\right\rangle,$ $|1\rangle \equiv
\left|\uparrow\right\rangle \}_i$ for a collection of $N$ individually
trapped single electrons $0 \le i \le N$.

\subsection{Planar Penning traps}

An application of trapped electrons for QC requires a clear route for the
miniaturization of single-electron-traps and a design for the multiple interconnections
between those individually  trapped electrons to allow for quantum gate operations. The
planar micro fabrication technology is well developed and helps designing and
fabricating a scalable interconnected multi-trap array. Our approach can be seen in
context with other proposals for single ions in linear \cite{DEN05} or two dimensional
\cite{CHIA08} arrays of Paul traps, which are possibly electrically coupled
\cite{HARTMUT}. The use of a Penning trap with a large planar ion crystal for QC has
been discussed in Ref.~\cite{POR06}.

\begin{figure} [htbp]
    \centerline{\epsfig{file=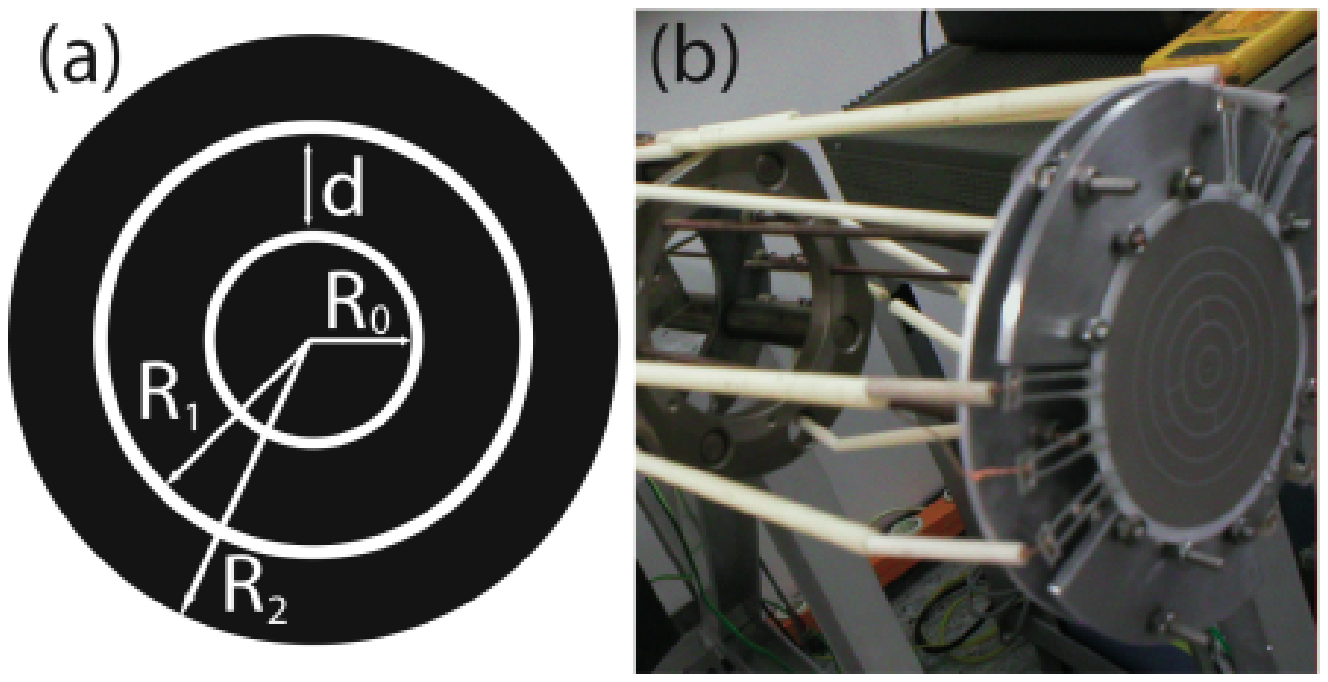,width=10cm}}
    \vspace*{13pt}
    \fcaption{\label{Fig: Fotoschema} (a) Sketch of a generic planar trap
    design consisting of a central disk of radius R$_0$ and a ring electrode
    of width d and outer radius R$_1$. A voltage can be applied between the
    two electrodes while the surface parts outside R$_2$ are held at ground
    potential. The electrode with radius R$_2$ is used to compensate
    anharmonicities of the potential. (b) Photograph of a prototype planar trap with several ring
    electrodes surrounding the central disk. The total diameter of the silver
    plated Al$_2$O$_3$ ceramic disk is D = 48~mm, with electrodes of
    R$_0$ = 2.5~mm, R$_1$ = 5.8~mm, R$_2$ = 9.1~mm and d = 3~mm.
    The trap is electrically connected and mounted such that it can be
    inserted into the magnetic field.}
\end{figure}

The basic building block of our approach is a single electron Penning trap consisting -
in its simplest form - of a central disk and a ring electrode on a planar surface, see
Fig.~\ref{Fig: Fotoschema}(a), where the magnetic field B$_0$ is perpendicular to the
plane. If we apply a voltage between these two electrodes and keep the surrounding
parts on ground, the electric potential in axial direction shows a minimum serving for
axial electron confinement.

{\em Harmonic axial potential: } The harmonicity of the trap potential is of importance
for a number of reasons: Particle trapping is more stable and non-linear resonances
\cite{ALHEIT1995} are avoided such that a long storage time is guaranteed. Second, the
qubit measurement relies on the detection of an inductive voltage pick-up at $\omega_z$
from the electron motion which is of the order of a few nV and is therefore detected after
a narrow-band filtering and amplification, see sect.~\ref{s:detection}. Therefore, the
signal processing would be difficult, if the motional frequency would depend on the
axial quantum state $k$. Thirdly, the cooling of axial motion, see
sect.~\ref{s:cooling}, requires fairly equidistant energy levels.

\begin{figure} [htbp]
    \centerline{\epsfig{file=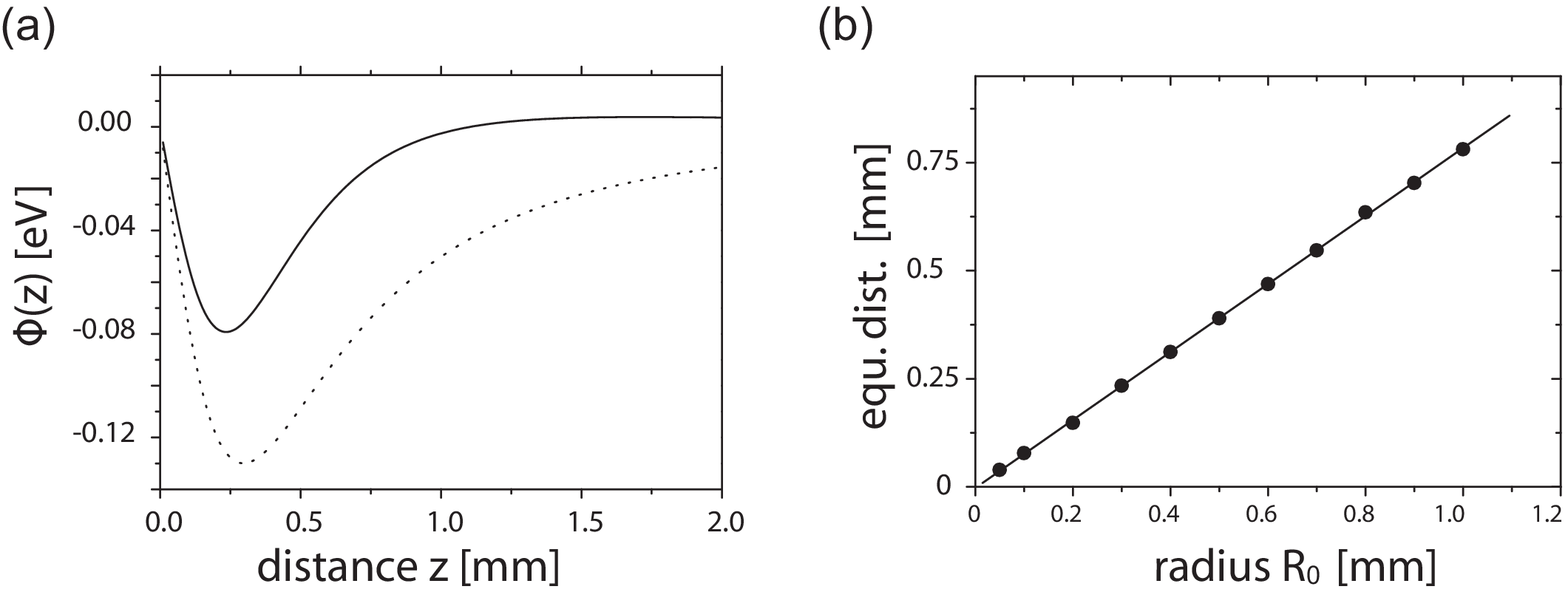,width=12cm}}
    \vspace*{13pt}
    \fcaption{\label{Fig: Pot} (a) Electric potential $\Phi^{el.}(z)$
    for a trap with R$_0$ = 300~$\mu$m, R$_1$ = 600~$\mu$m, R$_2$ = 900~$\mu$m
    and U$_0$ = 0~V, U$_1$ = +0.5~V, U$_2$ = 0~V leading to a potential with
    a minimum near z = 296~$\mu$m and an axial frequency of
    $\omega_z/(2\pi)$ = 89.9~MHz (dashed).  Anharmonicities are minimized
    by changing the compensation voltage to U$_2$ = -0.417~V. The minimum
    is shifted to a distance of 234~$\mu$m for the axial frequency of 
    99.0~MHz (solid line).
    (b) Position of the equilibrium position above the surface for the case
    of a compensated potential. The distance scales linearly as 0.78~R$_0$.}
\end{figure}

A particular feature of planar traps is the missing mirror symmetry around the electric
potential minimum which makes it impossible to create a perfect harmonic potential in
axial direction. However, as in the case of three-dimensional Penning traps additional
electrodes formed out of additional rings serve for partial compensation of
anharmonicities. Fig.~\ref{Fig: Pot}(a) shows the calculated potentials with and
without compensation. Even though the compensation of higher order coefficients in the
Fourier expansion of the potential decreases the depth of the trapping potential, the
voltages for trapping are easily increased. Note that with additional electrodes not
only the shape of the potential is changed but also the distance of the potential
minimum above the surface, see Fig.~\ref{Fig: Pot}(b). This will allow to investigate
in detail possible influences of surface effects on the coherence properties of the
trapped electrons.

{\em Experimental results from mm-sized planar traps: } Operation of two prototype
planar traps at room temperature with a total diameter of D = 48~mm and 20~mm with clouds
of hundreds of electrons shows the expected results \cite{GAL06}: After loading the
trap we can excite the different oscillations in axial and radial directions and find
agreement with the calculated frequencies. For all details about the operation of this
electron trap, the detection of electrons in this setup, and the excitation of motional
frequencies of the trapped electrons please see Ref.~\cite{GAL06}. A planar trap with
D = 20~mm has been operated at T = 100~mK. Clouds of electrons have been stored and
detected for 3.5~h, the observed axial frequencies are well understood from calculated
potential shapes such as in Fig.~\ref{Fig: Pot}(a) \cite{BUSHEV2008}.

\begin{figure} [htbp]
    \centerline{\epsfig{file=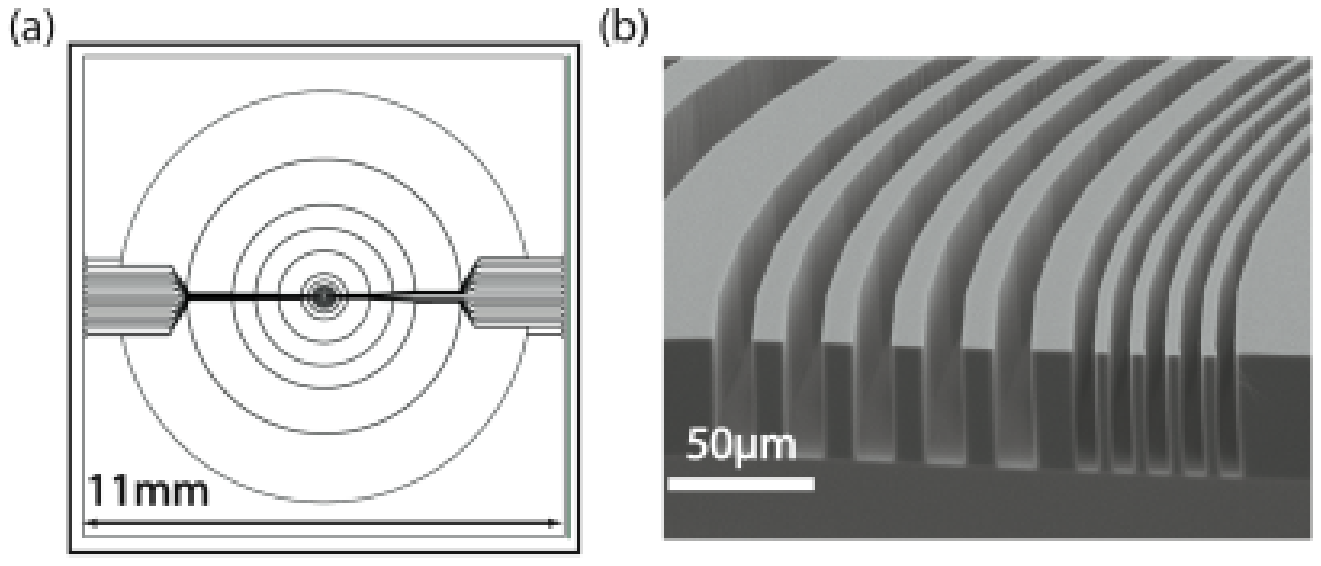,width=12cm}}
    \vspace*{13pt}
    \fcaption{\label{Fig: onion} (a) Design of a planar concentric
    electrode structure for a micro trap with center electrode radius
    that can be varied for a small trap configuration with
    R$_{0/1/2}$ = 50/100/150~$\mu$m, intermediate sizes up to a large
    trap size with R$_{0/1/2}$ = 1500/3000/4500~$\mu$m. (b) Electron
    scanning microscope image of a test photo resist structure with
    aspect ratio of 5~$\mu$m : 30~$\mu$m before casting with a gold
    surface.}
\end{figure}

{\em Micro fabricated single electron trap with multiple ring electrodes: } The design
of a micro fabricated single electron trap is shown in Fig.~\ref{Fig: onion}(a).
Concentrically arranged ring electrodes $E_{1..N}$ around a circular central electrode
$E_0$ can be operated as an {\em effective} circular center electrode with $U(E_0,
E_1,..., E_i)=U_0$ being surrounded by an {\em effective} ring electrode with
$U(E_{i+1}, E_{i+2},..., E_N)=U_1$ such that the initial trapping volume of the Penning
trap is large. After loading the trap using a deep and wide potential it might be
modified adiabatically to reach stiff and harmonic conditions with $U(E_0)=U_0$,
$U(E_1)=U_1$, and $U(E_2)=U_2$ serving for the compensation. The distance of the
electron to the electrode surface is adjusted by operating the central electrode
together with a certain number of rings $E_1, ..., E_i$ on the same voltage such that an
effective radius R$_0$ and hereby the potential minimum is controlled, see
Fig.~\ref{Fig: Pot}(b). We refer to the sect.~\ref{s:decoherence} for a detailed
discussion of surface decoherence effects that might be investigated by
changing the electron-surface distance in a well determined way.

\begin{figure} [htbp]
    \centerline{\epsfig{file=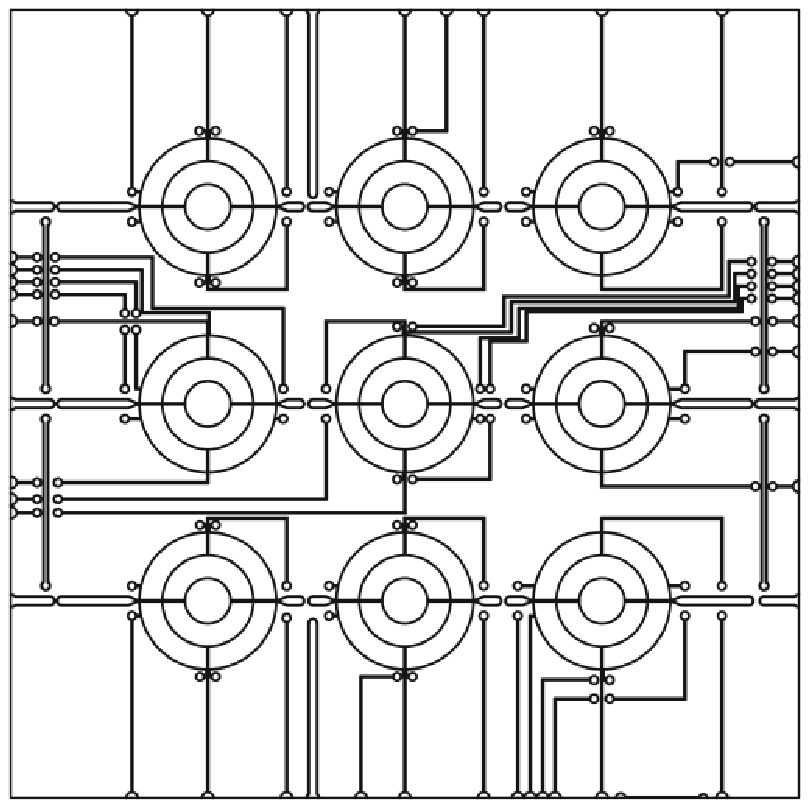,width=12cm}}
    \vspace*{13pt}
    \fcaption{\label{Fig: quprom} Planar electrode structure
    for the arrangement of nine interconnected micro traps of
    the standard type each with R$_0$ = 300~$\mu$m, R$_1$ = 600~$\mu$m
    and R$_2$ = 900~$\mu$m. The dimensions correspond to the
    calculated potential as plotted in Fig.~\ref{Fig: Pot}(a).
    Bonding pads on the gold surface allow for a flexible
    variation of the connections.}
\end{figure}

\section{Planar electron trap arrays}
For a scalable electron quantum processor the coupling of electrons is important, each
of them in its single trap. This coupling is provided by Coulomb interactions, either
direct or mediated by wire connections between the micro traps in the array. The
corresponding design is shown in Fig.~\ref{Fig: quprom}. It allows connecting the
central circular electrode of one of the traps electrically to another one such that
the axial motion of electrons might be coupled. With the chip design, the number of
mutually coupled traps can be chosen with the gold bonding on the chip such that a
number of interconnected qubits is established. Even with a normal conducting wire,
coherence is preserved and quantum gates are feasible as discussed in
sect.~\ref{s:wiregate}.

\begin{figure} [htbp]
    \centerline{\epsfig{file=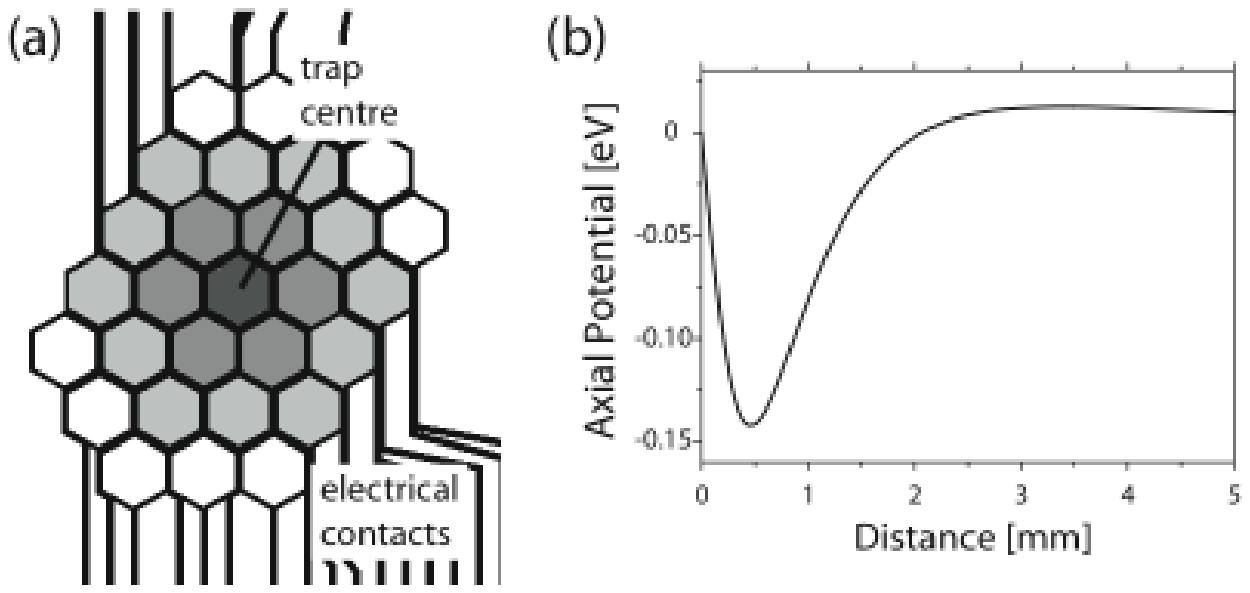,width=12cm}}
    \vspace*{13pt}
    \fcaption{\label{Fig: honey} (a) Sketch of the array of
    29 planar electrodes, each with a dimension of 0.43~mm,
    such that all electrodes be addressed by a different voltage.
    A trap might be formed above one of the hexagons at U$_0$=0
    (dark grey in the figure) which is surrounded by six hexagons
    at U$_1$=+0.6~V which are again surrounded by twelve hexagons
    at U$_2$=-0.5015~V for the compensation of axial anharmonicity.
    All other electrodes are at U=0~V. (b) The axial potential above
    the trap center is obtained by a numerical simulation.
    We find a trap depth of 0.14~eV in the harmonically compensated
    potential. A single electron would be trapped at a distance of
    0.46~mm with an axial frequency of $\omega_z/(2\pi)$= 59.5~MHz.}
\end{figure}

The second approach consists of a planar structure where the voltages on hexagon shaped
electrodes (Fig.~\ref{Fig: honey}) are controlled from outside such that the lateral
and vertical position of the potential minimum of each trap is adjustable. The
hexagonal electrode structure allows a freely programmable number of axial potentials.
The mutual distance of electrons confined in different potential minima can be reduced
such that the direct Coulomb interaction serves for quantum gates. We simulated the
potential for a single trap using a self-developed boundary element method
\cite{SCH06}. The voltage U$_2$ for the twelve surrounding hexagons was 
found that optimizes the harmonicity of the potential, see Fig.~\ref{Fig: honey}(b).

A group at Imperial College London has proposed recently yet another design for planar
traps using wires as electrodes \cite{CAS05} and to transport electrons between sites.

\section{Overall experimental apparatus}

Electrons are loaded from a field emission cathode \cite{STA05} into the array of
planar traps. In general the number of electrons confined in a single planar trap will
be larger than one. Excess electrons therefore may be detected and then
removed from the trap through the excitation of their axial oscillation by a
radio-frequency field, applied to the trap electrodes. The assembly of traps is 
located at the center of a superconductive solenoid of $B_0 \sim 
3\,{\rm T}$ magnetic field
strength with the direction perpendicular to the planar surface. The trap is housed in
a vacuum container in thermal contact to a dilution refrigerator\footnote{Oxford Inc.,
Kelvinox 100}. At a temperature below 100~mK a vacuum well below 10$^{-11}$~mbar is
ensured by cryogenic pumping. The low residual background pressure excludes
perturbations by collisions of the electron with residual gas molecules or He atoms for
a sufficiently long time. In thermal equilibrium with the cold environment the
electrons cool into the quantum mechanical ground state cyclotron motion by energy loss
from synchrotron radiation \cite{PEI99}.

The planar trap is equipped with a pick-up coil\footnote{Typically 
connected to the innermost circular electrode.}\, 
for the axial motional frequency, followed by an amplifier which
sends this rf-signal to the outside of the cryostat. A rf-signal near $\omega_z$ can
also be fed-in from an external frequency source for an excitation of the electron.
Microwave guides allow sending in frequencies near $\omega_{L}$ or $\omega_+$ serving
for the diagnosis and for qubit gate operations. Sect.~\ref{s:gate} will describe the
required microwave pulses for universal single and two-qubit operations.

\section{Electron spin qubit detection}
\label{s:detection}

In our setup electrons are detected non-destructively by the induced image charges of
the axial motion. The detection provides a direct measurement of
\begin{description}
\item[a)] the presence of a single electron or the number of trapped electrons and
\item[b)] the energy of the stored electrons in the axial degree of freedom.
\end{description}
If an inhomogeneous magnetic field B$_{2}$ is superimposed to the quantization field
B$_0$ the ``continuous" Stern-Gerlach effect \cite{BRO86,DYC93,DIE98,DEH86} allows to
\begin{description}
\item[c)] detect the spin state $\{\left|\uparrow\right\rangle, \left|\downarrow\right\rangle \}$ of the electron and also
\item[d)] the cyclotron motion.
\end{description}
The technically most demanding requirement is a low-noise cryogenic amplification of
the ultra-weak inductive pick-up by using a preamplifier at T=0.1~K followed by an
amplifier at T=0.5~K \cite{PEIL99}. The output signal is detected by a low-noise
spectrum analyzer\footnote{Signature B, Anritsu Inc., with a displayed average noise
floor of -167 dBm}.

{\em Axial motion detection schemes: } To understand the detection method used for (a)
and (b), we imagine that the electric LC-circuit composed by the 
pick-up coil (a helical copper
resonator) and the capacitances from the trap electrodes shows up on the spectrum
analyzer by the enhancement of noise within a bandwidth that corresponds to the Q-value
of the resonance. The values of L and C are chosen such that the resonance is centered
at $\omega_{z}$. Initially, the electrons motional energy in the axial degree of
freedom is high and the noise measured by spectrum analyzer exceeds the noise
background of the naked LC-circuit. The coupling to the axial circuit serves for
cooling the axial degree of freedom to the environmental temperature (``resistive
cooling"). For low electron energy, a dip appears in the noise spectrum, as the 
motion of the trapped particle ``short-cuts" the LC-circuit exactly at the axial trap
frequency \cite{DIE98}.

The {\em Stern Gerlach effect } is understood here as the variation of the axial frequency
conditioned on the electron spin state~\cite{DEH86,WER02}. Any inhomogeneous magnetic field B$_2 \propto
z^2$ superimposed to B$_0$ leads to a coupling of the spin and motional modes of the
trapped electron. This inhomogeneity is generated by a ring of ferromagnetic material.
The size of the axial frequency shift is controlled by the magnitude of the
inhomogeneity, typically a fraction of $10^{-6}$ of $\omega_{z}$. As we measure the
frequency of the axial dip we can deduce the spin state $\{\downarrow, \uparrow \}$.
The method has been applied for the measurement of the magnetic moment of the free and
bound electron \cite{DYC93,ODO06,WER02}. In our proposal, the Stern Gerlach effect can
be dynamically controlled even with constant magnetic fields 
B$_{0,2}$ by varying the
distance of the electron to the surface: the detection of the spin direction
should happen at a location close to the central circular trap 
electrode, e.g., where the
inhomogeneity is comparatively large, and the pick-off signal for axial detection is
relatively high. Further away from the surface, in a more homogeneous field region,
the spin state remains undisturbed. The controlled variation of all electron
positions also allows for a direct and unambiguous addressability of those qubits by 
defining their individual Larmor and reduced cyclotron frequencies. For the axial motion the
addressing can also be achieved by a slight detuning of the trapping potentials,
resulting in different trapping frequencies for every individual electron.

\subsection{Sideband cooling and axialisation of the electron orbit \label{s:cooling}}

To initialize the electrons for quantum gate operations, cooling techniques are
employed. As shown by Gabrielse and coworkers \cite{PEI99}, the cyclotron energy is quantized
and the electron cools energy by synchrotron radiation to the cryogenic 
environment, ultimately reaching the ground state $|n=0\rangle_c$
where it remains if $k_B T \ll \hbar \omega_c$. Microwave pulses coupling the
axial and cyclotron motion $|n=0\rangle_c|n\rangle_{z} \rightarrow
|n=1\rangle_c|n-1\rangle_{z}$ followed by spontaneous decay $|n=1\rangle_c \rightarrow
|n=0\rangle_c$ will lead to axial cooling \cite{COR90,VER02}.

The axialization of the electron motion uses a resonantly driven excitation at the
magnetron frequency such that the orbit is stabilized
\cite{SCHWEIKHARDT93,DEH86,POVELL2002}.

\section{Electron qubit gates}
\subsection{Single qubit gates}
\label{s:gate}

A quite natural choice to encode qubits with trapped electrons is the
particle spin. Indeed, the two possible spin orientations $\{\left|\uparrow\right\rangle,
 \left|\downarrow\right\rangle\}$ in the trap magnetic field may represent a qubit. Alternatively,
the motional degrees of freedom $\{|0\rangle_c, |1\rangle_c\}$ or $\{|0\rangle_z,
|1\rangle_z\}$ might serve to encode qubits. Appropriate rf or microwave pulses drive
gate operations, addressed to single qubits, while all other qubits stay unperturbed as
their frequencies are shifted far from resonance. For the motional qubit states,
measures have to be taken not to populate Fock states larger than $|n=1\rangle$ by
composite pulse techniques \cite{GULDE2003,SCH03a,CHILDS2001,Stortini}, or resorting to
anharmonicities \cite{CIA01,CIA02,CIA04}.

\subsection{Two qubit gates}
The qubit bases mentioned above allow for various types of conditional qubit dynamics
\cite{CIA03,ZuritaSanchez2008}.
Here we restrict ourselves to scalable methods and
focus on a direct spin-spin interaction mediated by a magnetic field gradient and by an
electrical coupling of the electrons axial motion via a direct connection of trap
electrodes. While the first one has the advantage of working without ground state
cooling of the axial motion, the second one has the advantage of coupling even distant
electron sites in the array.

\subsubsection{Effective spin-spin interaction}
\label{spinspincoupling}

Consider an array of $N$ electrons that are exposed to an 
inhomogeneous magnetic field producing a linear gradient $B_{1} \propto b x$.
This means that the
free electron cyclotron frequency $\omega_c$, the spin Larmor frequency $\omega_L$, the
reduced cyclotron frequency $\omega_+$ and the magnetron frequency $\omega_-$ all
depend on the particle position in the array. For instance, with a magnetic gradient
of $50$~T/m, the spin flip frequencies of two electrons in neighboring traps,
separated by a distance of the order of 10$^{-3}$~m, differ from each other by few MHz.
This value, though small in comparison with the typical spin frequency $\omega_L
/(2\pi) \sim 100$~GHz, is enough to individually address the qubits via microwave
radiation. Moreover, the array of traps can accomodate up to tens of qubits, with their
frequencies spread over a range of 100~MHz.

Importantly, the linear magnetic gradient couples the motional degrees of freedom of
each particle \cite{WUNDERLICH2001} to its spin state:
\begin{equation}
  H_{spin}= \frac{g}{2} \, \mu_B \, \mbox{\boldmath ${\sigma}$ }
                  \cdot \mathbf{B(x,y,z)}         \label{H_spin}
\end{equation}
where $\mu_B$ denotes Bohr magneton and $\sigma_x$, $\sigma_y$, and
$\sigma_z$ are Pauli matrices.  The long-range Coulomb interaction
establishes an effective coupling
between the spin qubits.  As shown in
Ref.~\cite{ciaramicoli_pra_2005}, the spin-spin interaction reads:
\begin{equation} \label{H_spin_final}
 H'_{spin-spin} \simeq
       \frac{\hbar \pi}{2} \sum_{i>j}^N  J_{i,j} \hspace{2mm} \sigma_{z,i} \hspace{2mm}
             \sigma_{z,j}
\end{equation}
The spin-spin coupling strength $J_{i,j}$ is tunable, since it depends on external
parameters like the strength of the applied magnetic field gradient $b$, the inter-trap
distance $d_{i,j}$, and the characteristic trapping frequencies:
\begin{equation}
    J_{i,j} = \frac{1}{4\pi \epsilon_0} \frac{g^2 \mu_{B}^2 e^2}{2\pi \hbar m^2} 
              \frac{b^2}{\omega_z^4 d_{i,j}^3}
\end{equation}
The Hamiltonian Eq.~(\ref{H_spin_final}) is formally analogous to the Hamiltonian
describing a system of nuclear spins used to perform NMR quantum computation. This
similarity suggests that techniques developed and tested in NMR experiments can be
readily exported to trapped electrons in order to implement quantum logic gates.
We quote typical orders of magnitude for this spin-spin coupling in
Table~\ref{t:spin-spin-coupling}.

\begin{table}[hb]
\tcaption{\label{t:spin-spin-coupling}Coupling strength of the spin-spin interaction $J$ in units
$\omega/(2\pi)$ for different values of the magnetic field gradient $b$ and
inter-particle distance. The axial frequency $\omega_z/(2\pi)$ is 100~MHz.}
\centerline{\footnotesize\smalllineskip
\begin{tabular}{|c|c|c|c|}
\hline
        &100~$\mu$m & 50~$\mu$m & 10~$\mu$m \\
\hline
50~T/m  & 2.3~Hz    & 18~Hz     & 2300~Hz \\
500~T/m & 0.23~kHz  & 1.85~kHz  & 230~kHz \\
\hline
\end{tabular}}
\end{table}

\subsubsection{Coherent wire coupling}
\label{s:wiregate}

We consider next a two-qubit gate implemented with the axial oscillation of two
electrons in separate traps that are connected with an ``information exchange wire''.
This couples the oscillating image charges on the two electrodes to each other. The
case of a superconducting connection \cite{SCHOELLKOPF2007} was analyzed in
Ref.~\cite{Soerensen04a}. We point out here that even with a normally conducting wire,
a coherent trap coupling is possible despite the presence of thermal noise. The basic
method is a quantum theory of electric signal propagation and noise in resistive
circuits at finite temperature \cite{ZuritaSanchez06a}. Applying this theory to the two
coupled electrons gives (i) a rate $\Omega_{12}/(2\pi)$ for the coherent swapping of
excitations between the traps; (ii) dissipation and decoherence rates for both traps,
due to the thermal resistance of the wire; and (iii) a dissipative cross-coupling of
the electric field fluctuations due to the fact that both traps couple to the same
environment.

\begin{table}[]
\tcaption{\label{t:wire-coupling}Coupling strength of the wire
mediated interaction $\Omega_{12}$ in units $\omega/(2\pi)$ for
different trap radii $R_0$ and inter-trap distances $d$.  The axial
frequency $\omega_z/(2\pi)$ is 100~MHz.  According to the results
shown in Fig.~\ref{Fig: Pot}, $h$ is related to $R_0$ for a
harmonically compensated trap.  We assume capacitances $C_0 = R_0
\times 100\, {\rm fF/mm}$, $C_w = d \times 66\, {\rm fF/mm}$,
realistic for the trap layout of Figs.~\ref{Fig: onion} and \ref{Fig:
quprom}.
}
\centerline{\footnotesize\smalllineskip
\begin{tabular}{|r|c|c|c|c|c|}
\hline
&$d$ = 100~mm & 10~mm & 1mm & 100~$\mu$m & 10~$\mu$m\\
\hline
$R_0$ = 1mm & 0.12~Hz & 1.0~Hz & 3.2~Hz & -- & -- \\
10$\mu$m & 1.3~kHz & 13~kHz & 130~kHz & 990~kHz & 3\,200~kHz\\
\hline
\end{tabular}}
\end{table}

In the low-frequency regime the wire mainly acts as a capacitive coupler. The coherent
coupling is described by the Hamiltonian
\begin{equation}
    H_{z,12} = \hbar \Omega_{12} \left(
    a_{z,1}^\dag  \hspace{1mm} a_{z,2}^{\phantom\dag}
    +
    a_{z,1}^{\phantom\dag} \hspace{1mm} a_{z,2}^\dag
    \right)
    \label{eq:two-axial-coupling}
\end{equation}
where $a$ and $a^\dagger$ denote the creation and annihilation of axial
quanta, and the
Rabi frequency~\cite{ZuritaSanchez2008}
\begin{equation}
\Omega_{12} = \frac{e^2}{2 m \omega_{z}} \frac{R_{0} / (h^2 +
R_0^2)^3}{2 C_0 + C_\mathrm{w}} \label{eq:coherent-coupling}
\end{equation}
depends on height of the electron above the trap electrode $h$ and
the center electrode radius $R_0$.
The capacitance $C_0 \sim \pi \varepsilon_0 R_0$ describes
the intrinsic electrode capacitance and $C_\mathrm{w}$ is the capacitance
of the wire connecting the two traps. The shorter the wire ($C_\mathrm{w}$ decreases)
and the smaller the trap electrodes ($R_0$ and $C_0$ decrease), the stronger the
coherent coupling \cite{ZuritaSanchez2008}. Typical values are quoted
in Table~\ref{t:wire-coupling}.

\section{Non-perfect quantum operations}
\label{s:decoherence}

We identify two classes of imperfections which make the trapped electrons loose their
ability to persist in certain quantum superposition states. An obvious source of
dephasing are fluctuations in the control parameters for a quantum operation such as
the trap dc-voltages, magnetic field stability or microwave frequency and phase. Table
\ref{t:errors} lists several of the effects for typical parameters in the experiment. A
second class of decoherence arises from more fundamental couplings where the
fluctuations are of quantum or thermal origin, typical examples are the spontaneous
decay of the cyclotron Fock states or the Johnson noise in a metallic conducting
surface which couples to the axial motional degree of the electron.

\subsection{Fluctuating control parameters}
\label{s:bad-control}

The noise from fluctuations of the \emph{trap control voltages} has
been estimated as follows: the 24bit voltage supply allows a stability
of $\Delta U/U \simeq 6 \cdot 10^{-8}$. This translates into
(quasi-static) fluctuations of the axial trapping frequency, $\Delta \omega_{z} /
\omega_{z} = \frac12 \Delta U/U$. A characteristic dephasing time
$\tau_{z}$ for superposition states of the axial qubit can be
defined by $\Delta \omega_{z} \tau_{z} \sim 1$. For the axial
frequency of 100~MHz mentioned above, this gives $\tau_{z} \sim 50\,{\rm
ms}$. These fluctuations can be compared to dynamical noise that
leads to a diffusive increase of the relative phase at the rate
$\gamma_{z} \sim (\omega_{z}/2)^2 S_{U} / U^2$ where $S_{U}$ is
the power spectral density of the voltage noise.
(We normalize $S_{U}$ such that the rms voltage noise $\delta
U$ in a frequency band $\Delta\nu$ is $(2\,S_{U}\,\Delta\nu)^{1/2}$.)
The same reasoning applies to the dephasing of the cyclotron and spin
qubits where the relative phase of superposition states is randomized
by magnetic field fluctuations.
Values that we estimate as realistic for
the cryogenic electronics used in our experiments are collected in
Table~\ref{t:errors} and show that the limiting factors are DC electric
field stability (for the axial qubit) and magnetic field fluctuations
(for the cyclotron qubit).

To obtain a well-defined control of the qubit states, external rf and
microwave fields are required and their amplitude and phase 
uncertainties have to be considered.
Fortunately one benefits here from the cryogenic environment and the
fact that the excitation signals can be created by non-linear effects
directly in the cold region.
Hence in the ``off'' state, when the drives are not active, the 
externally introduced additional noise background is negligible.
The remaining noise from the trap environment can be estimated
as described below in sect. \ref{s:bad-environment}, concerning the cyclotron and spin
transitions.
When the drives for manipulating the quantum states (axial, cyclotron
or spin) are turned ``on'', their amplitude and phase uncertainties 
within the respective excitation times come into play.
It is expected that the respective rf and microwave signals, while
guided from the room temperature region into the cryogenic trap setup,
do not suffer considerable short term amplitude ot phase variations.
Therefore, the same considerations as for comparable room temperature
setups and the manipulation of qubits in trapped ions can be applied
\cite{GULDE2003,LEI03,MAJ06}.
However, taking into account manipulation rates of qubit states not
faster than a few kHz, it can be expected that the latter effects are
rather small compared to all other decoherence mechanisms.

\begin{table}[hb]
\tcaption{\label{t:errors}Error sources for single qubit and two qubit
interactions.  Axial qubit frequency 100\,MHz, cyclotron and spin
qubit frequency 100\,GHz.  The axial qubits are coupled by a gold wire
with specifications mentioned after Eq.(\ref{eq:figure-of-merit}).}
\smallskip
\centerline{\footnotesize\smalllineskip
\begin{tabular}{|c|c|c|c|c|}
\hline &
\multicolumn{2}{c|}{electric field} & magnetic field & axial wire coupling \\
\hline
relative fluctuation
&
$6 \cdot 10^{-8}$ (DC)         & $10^{-11} / \sqrt{ {\rm Hz}Ê}$ @ 1\,kHz
& $10^{-11} / \sqrt{ {\rm Hz}Ê}$ @ 1\,kHz
& Johnson noise @ 100 mK \\
decoherence rate
& $\sim 10\,{\rm s}^{-1}$       &   $\sim 10^{-5}\,{\rm s}^{-1}$
& $\sim 10\,{\rm s}^{-1}$
& $10^{-2}\, {\rm s}^{-1}$\\
\hline
\end{tabular}}
\end{table}

\subsection{Thermal and quantum noise from the trap environment}
\label{s:bad-environment}

The \emph{cyclotron oscillation} is driven towards its ground state
via electric dipole coupling to electric vacuum fluctuations.  This
happens, in free space, at the spontaneous emission rate
\begin{equation}
\gamma_c = \frac{ e^2 \omega_+^2 }{ 3\pi\varepsilon_0 m c^3 }
\label{eq:spont-emission}
\end{equation}
corresponding to a decay time of a few seconds. Thermal fluctuations
can be ignored since at 100\,mK, the occupation numbers of states with
one or more photons (energy $\hbar\omega_{+}$) are negligible.
In our apparatus with a half-open geometry, standing waves may form
(cyclotron wavelength of $2.8\,{\rm mm}$ at $100\,{\rm GHz}$) such
that the spontaneous emission rate is slightly modified \cite{BRO86}.

The \emph{spin state} of the trapped electrons suffers heating and
decoherence from magnetic field fluctuations.
If we follow the master equation approach outlined in Ref.~\cite{CIA04},
the dynamics of the relevant matrix elements of the spin density operator
$\rho^{\mathrm{spin}}$ is given by
\begin{eqnarray}
\dot{\rho}_{\downarrow\downarrow}  &=& -2 \Gamma_- \rho_{\downarrow\downarrow}
                     +2 \Gamma_+ \rho_{\uparrow\uparrow} , \\
\dot{\rho}_{\downarrow\uparrow} &=& {\rm i} \omega_{L} \rho_{\downarrow\uparrow}
- 3 (\Gamma_- + \Gamma_+) \rho_{\downarrow\uparrow} .
\end{eqnarray}
Unintended spin flips $\left|\downarrow\right\rangle \rightarrow
\left|\uparrow\right\rangle$ ($\left|\uparrow\right\rangle \rightarrow \left|\downarrow\right\rangle$)
take place with a rate $2 \Gamma_-$ ($2 \Gamma_+$), while the
coherence decays with a rate $3 (\Gamma_- + \Gamma_+)$.  The rates
\begin{equation}
  \Gamma_{\pm} \equiv \left(
                 \frac{g e}{4 m}
              \right)^2 S_B(\pm \omega_L)
\end{equation}
depend on the spectral density $S_B(\omega)$ of the magnetic field
fluctuations at the Larmor frequency,
\begin{equation}
    S_{B}( \omega ) =
    \int\limits_{-\infty}^{\infty}\!{\rm d}t\,{\rm e}^{ - {\rm i}
    \omega t }
    \langle B_{z}( t ) B_{z}( 0 ) \rangle
    \label{eq:def-B-spectrum}
\end{equation}
In the strong magnetic fields and low temperatures considered here,
this spectrum is dominated by vacuum
fluctuations, as for the cyclotron oscillation, and modified by the
trap geometry. Relative to the cyclotron decay
rate~(\ref{eq:spont-emission}), spin decay is suppressed by a factor
$\hbar\omega_{L} / m c^2 \sim 10^{-9}$ and hence negligible. Low
frequency fluctuations make the phase of $\rho_{\downarrow\uparrow}$
drift, limiting the phase coherence as discussed in
Subsection~\ref{s:bad-control}.

The \emph{axial qubit} has a level spacing small compared with the environmental energy $\hbar \omega_z
\ll k_B T$ such that this oscillator ends in a thermal state $n_z \gg$ 1. However, the
electron can be driven to the axial ground state using sideband-cooling techniques as
mentioned in sect.~\ref{s:cooling}. Once the axial oscillator is prepared in its
vibrational ground state, the coupling to quasi-static electric field fluctuations from
charges and currents in the circuitry surrounding the trap becomes important. The
relevant quantity is the spectral power $S_{E}( \omega_{z} )$ of the electric field
fluctuations $E_{z}$ at the axial frequency, defined by analogy to
Eq.(\ref{eq:def-B-spectrum}).
This leads
to a heating rate:
\begin{equation}
    \gamma_{z} = \frac{ e^2 }{ 2 \hbar m \omega_{z} }
    S_{E}( \omega_{z} )
    \label{eq:heating-rate-vs-spectrum}
\end{equation}
Heating also destroys superposition states and limits the coherence
time for the axial qubit.
Electric noise is generated from several sources, such as patch charge
fluctuations and thermal Johnson noise from the detection circuit. Noise inherent to the
electrode material turns out to amount for a negligible heating rate.
Additionally, the
fluctuations of the trap control voltages contribute significantly to the
axial dephasing rate. They have been estimated in
Subsection~\ref{s:bad-control} above.

The heating due to \textit{patch charge fluctuations} has been observed in
radio-frequency ion traps with sub-mm sizes \cite{SCHULZ2008,Epstein07}. The heating
rate decreases strongly for a temperature of 4~K \cite{Labaziewicz07}. Compared to the
Paul traps for ions, our experiment is operated at a temperature which is by two orders
of magnitude lower and the static dc-voltages for the electron confinement are not
driving the motion of patches. Therefore we expect a very low heating and decoherence
rate from this noise source. On the other side, the experiment with trapped electrons
may help understanding the physical mechanism behind.

Even if the electrode surfaces are prepared in an ideal way that a purely metallic
conducting surface without any surface contaminations is facing the trapped
electrons, the \emph{intrinsic damping properties} of the material will lead to noise.
The corresponding heating rate increases with ac-resistivity of the electrode material,
and the electrode temperature, and decreases with the axial frequency and distance $h$.
For gold electrodes with a radius of 100~$\mu$m we expect a rate of $\gamma_z \simeq
10^{-3} s^{-1}$ \cite{Henkel99c}.

\emph{Johnson noise from attached circuitry} plays a role only during the instance of
qubit measurement. During single qubit gate operations, the electrons are decoupled
from the circuit. If we work with the direct spin-spin coupling two qubit interaction,
see sect.~\ref{spinspincoupling}, also no circuit is attached and the gate is not
affected by this noise source. If for some reason, the measurement circuit needs to be
constantly kept resonant to the electron axial motion, any resistive element in this
circuitry generates Johnson-Nyquist voltage noise with a spectrum $S_V(\omega)$ = 2
$k_B T R_\mathrm{c}(\omega)$. This expression applies in the low frequency regime,
$\hbar\omega_z \ll k_B T$. The voltage noise is filtered by the circuit, in particular
capacitances, before arriving at the trap electrode \cite{Deslauriers06a}. For a
temperature on 100~mK, a circuit impedance of 100~k$\Omega$, a quality factor of
10${^3}$ and a trap central electrode radius of 100~$\mu$m we estimate $\gamma_{z}$
$\simeq$ 10~s$^{-1}$ (see Table~\ref{t:errors}).

Johnson voltage noise also affects the wire coupling of two qubits. Here the dissipation
stems from the finite resistance of the coupling wire. To illustrate the impact of
decoherence on the two-qubit wire gate, we consider the SWAP operation. The axial
qubits are initialized in the state $|n_{1} = 1, n_{2} = 0\rangle$, and the coherent
coupling via the wire is switched on for a time $\tau = \pi / \Omega_{12}$. After this
time, the excitation of qubit 1 should have been transferred to qubit 2 resulting in
the state $|n_{1} = 0, n_{2} = 1\rangle$. Due to the thermal fluctuations in the wire,
this state is only reached with a reduced fidelity. The relevant figure of merit for
this two-electron scheme is the maximum number of gate operations $N_\mathrm{max}$ that
are possible within a decoherence time~\cite{ZuritaSanchez2008}:
\begin{equation}
    N_\mathrm{max} \approx \frac{ \Omega_{12} }{ \pi \gamma_{z} }
    \approx \frac{ 3 \hbar }{ 4 \pi k_B T R_\mathrm{w}
    (2 C_{0} + C_\mathrm{w}) }.
    \label{eq:figure-of-merit}
\end{equation}
With parameters, as found for a gold wire of 20~$\mu$m diameter and
150~$\mu$m length, $C_0 \simeq C_W \simeq$ 10~fF, $R_W \simeq1\,{\rm
m\Omega}$ are expected to be realistic for our proposal, and we get
$N_\mathrm{max} > 10^5$ at 100~mK.

\section{Conclusion}

Certainly the various experimental techniques for QC show different assets and
difficulties. At the current status of research we summarize the assets of our proposal
along the criteria as formulated by DeVincenzo \cite{DIVINCENCO2001}:

Absolutely mandatory is the scalability of the approach that should at least allow
scaling to order 100 qubits. The dimensionality of the technique will determine how
efficient algorithms may be executed. Examples are the 1-dim. techniques such as linear
cold ion crystals, but also 2-dim. geometries which are for example realized in
circuits of superconducting elements on a planar substrate. An example of a 3-dim.
system is an optical lattice formed by three laser fields and holding cold atoms at the
lattice sites. Our electron trap array combines a 2-dim. geometry with individual
control over every qubit site. The size of micro Penning traps will allow a high number
of qubit sites on a single chip.

Coherence times are expected to be very long, due to the reduced noise in a cryogenic
and well shielded environment. The purely static control voltages improve further the
coherence properties.

Gate operation times for single and two qubit operations should be much shorter than
the decoherence time. According to our estimations this is largely fulfilled. All gate
operations can be implemented with standard radio frequency and microwave equipment, which
commercially delivers low phase noise. The gates which we described can be operated
highly parallel, e.g. the electrically conducting wire may serve as a quantum bus
allowing interactions between any pair of qubits in the register, not only between two nearest
neighbors.

In our quantum register, qubits are read-out with the help of the Stern Gerlach effect
which has been already used for decades in precision spectroscopy with trapped
electrons. The scheme allows even for a parallel read-out. For the initialization of
spin state we rely on a non-destructive measurement, while the motional qubits degrees
are laserless cooled.

Finally, the future QC may benefit from a combination and connectivity of different
techniques, in a similar way as our conventional computer uses a magnetic storage
device and an electronic processor from semiconductor materials, both connected by
normal conducting metal leads. Therefore it is of high importance that qubit
information may be transferred via electrically conducting wires. The low temperature
surface forming the single electron Penning traps may allow for an integration of and
coupling to superconducting circuit elements for QC.

In view of realizing quantum phase transitions one might think of grouping the qubit
sites, e.g. building linear or ring structures with either even or odd numbers of
sites, forming two dimensional lattices with or without frustration or configurations
of the Kagome type \cite{SANTOS95}. This would allow simulating interacting spin
systems with highly variable and designable properties.

\vspace{2mm}

We acknowledge financial support by the European Union within the
sixth framework programme (contract no.\ FP6-003772).

\nonumsection{References}

\bibliographystyle{mybibstyle}
\bibliography{quele08}

\end{document}